\documentclass[aps,physrev,preprint,groupedaddress,nobibnotes,superscriptaddress]{revtex4-2}

\usepackage{graphicx}
\usepackage{amsmath}
\usepackage{amssymb}
\usepackage{mathtools}
\usepackage[english]{babel}
\usepackage{siunitx}
\usepackage[colorlinks,allcolors={blue}]{hyperref}
\addto\captionsenglish{\renewcommand{\figurename}{Fig.}}

\usepackage[parfill]{parskip}
\begin{document}
\raggedbottom

\title{Deep learning enables large-scale inverse design of free-form metasurfaces}

\author{Viktor A. Lilja}
\affiliation{Department of Physics and Astronomy, Chalmers University of Technology, SE-412 96 Gothenburg, Sweden}

\author{Khosro Zangeneh Kamali}
\affiliation{Department of Physics and Astronomy, Chalmers University of Technology, SE-412 96 Gothenburg, Sweden}
\affiliation{Department of Physics and London Centre for Nanotechnology, King’s College London, London, UK}

\author{Mikael Käll}
\author{Philippe Tassin}
\affiliation{Department of Physics and Astronomy, Chalmers University of Technology, SE-412 96 Gothenburg, Sweden}

%\date{\today}

\maketitle

\section*{Abstract}
Metasurfaces are ultrathin optical elements that control light through carefully engineered structures smaller than the wavelength of light, enabling compact devices with functionalities that are difficult to achieve with conventional optics.
However, exploiting their full design freedom over large areas has been prohibited by the large computational cost of repeated electromagnetic simulations required for inverse design.
Here we introduce a physics-tailored deep-learning surrogate model that exploits the locality and symmetries of electromagnetic interactions to generalize from small-scale simulations to metasurfaces orders of magnitude larger.
The model captures nonlocal interactions while predicting optical fields more than four orders of magnitude faster than the conventional electromagnetic solver it is trained on, enabling gradient-based topology optimization under arbitrary illumination. 
We validate the approach through the inverse design and experimental realization of free-form metagratings with diffraction angles up to 75°, as well as a millimeter-scale free-form holographic metasurface more than 1,000 wavelengths across and comprising 1.6 billion degrees of freedom. 
By decoupling the scale of electromagnetic simulation from the scale of inverse design, our approach enables system-scale free-form photonic devices that retain nanoscale design freedom.

\section*{Main}

Metasurfaces provide a versatile platform for controlling light with subwavelength resolution, enabling compact optical components for imaging, sensing, wavefront shaping, optical computing, and more \cite{yu_flat_2014, khorasaninejad_metalenses_2017, zeng_performance_2025}. 
Dielectric metasurfaces are conventionally designed by arranging weakly interacting parameterized meta-atoms, such as nanopillars or nanofins, in a regular lattice.
While sufficient for simple wavefront manipulation tasks, this approach limits the design space to a small predefined set of geometries, thereby restricting the achievable device performance. The next generation of high-performance multifunctional metadevices will require design methods that can utilize the full range of attainable responses.

Topology optimization offers a route to go beyond the limits of conventional design by iteratively optimizing the material distribution of the metasurface, allowing arbitrary patterns to emerge \cite{jensen_topology_2011, kuznetsov_roadmap_2024, schul_inverse_2026, sell_large-angle_2017, phan_high-efficiency_2019, liu_topology_2025}.
This free-form inverse design approach has already enabled devices that outperform conventional metasurfaces \cite{sell_large-angle_2017, phan_high-efficiency_2019, sun_scalable_2025, liu_topology_2025}.
However, topology optimization requires computationally expensive repeated full-wave simulations of Maxwell’s equations, and its application has therefore been limited to small, periodic, or one-dimensional geometries \cite{choi_realization_2024, kang_large-scale_2024}.
Despite advances in highly optimized parallel solvers \cite{hughes_perspective_2021}, domain decomposition strategies \cite{lin_topology_2019, lin_overlapping_2019}, and symmetry-based simplifications \cite{sun_scalable_2025}, free-form optimization of two-dimensional metasurfaces beyond a few tens of wavelengths across for arbitrary objectives has not yet been demonstrated \cite{kuznetsov_roadmap_2024, choi_realization_2024, kang_large-scale_2024, schul_inverse_2026}.

Deep learning has emerged as a promising route to accelerate inverse design of electromagnetic devices by replacing costly full-wave simulations with fast neural network surrogate models \cite{nadell_deep_2019, an_deep_2019, hegde_deep_2020, wiecha_deep_2021, so_deep_2020, khaireh-walieh_newcomers_2023, augenstein_neural_2023, choi_realization_2024, kuznetsov_roadmap_2024}. 
However, for metasurfaces, the combination of free-form parameterization and large area results in an extremely high-dimensional design space that would require prohibitively large amounts of data for standard neural network architectures to learn.
Previous surrogate models for metasurfaces therefore typically rely on simplifying assumptions, such as neglecting nonlocal interactions (meta-atom coupling) \cite{gahlmann_deep_2022, choi_realization_2024, furat_physics-informed_2025}, allowing only a few parameters per meta-atom \cite{majorel_deep_2022, an_deep_2022, chen_artificial_2022, khoram_graph_2023, wu_general_2025, xu_deep-learning-enabled_2025, grand_design_2026}, or limiting the design space to one dimension \cite{lim_maxwellnet_2022, zhelyeznyakov_large_2023}.

\begin{figure*}[hbt!]
    \centering
    \includegraphics[width=\textwidth]{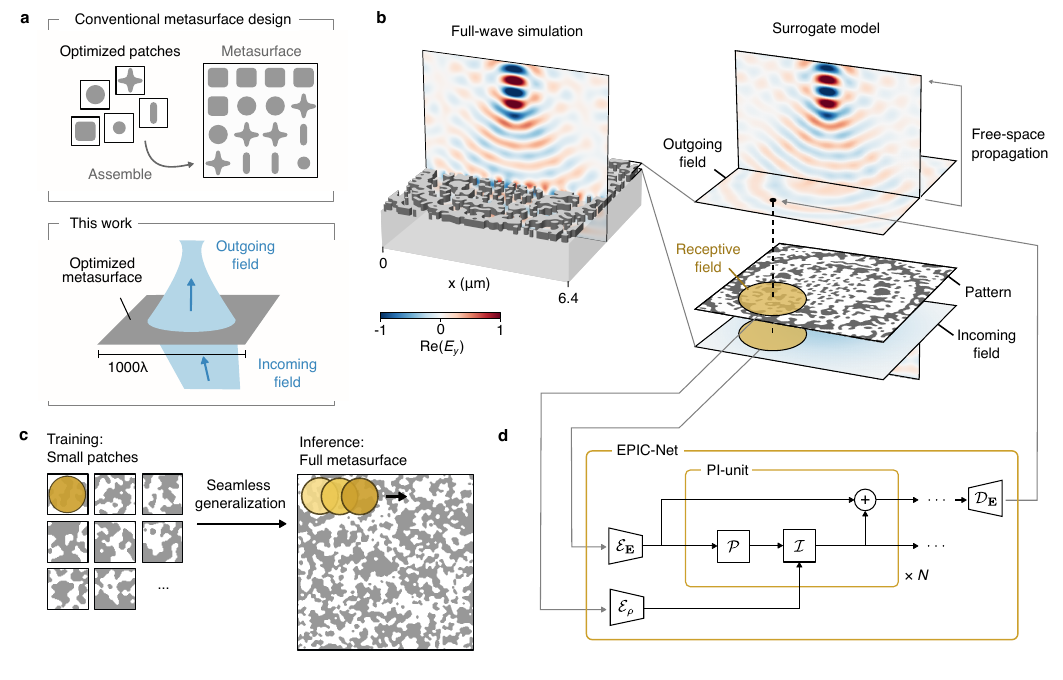}
    \caption{
    Large-scale metasurface simulation enabled by the EPIC-Net.
    (a) Comparison between conventional metasurface design and the full-metasurface optimization approach unlocked by this work.
    (b) Comparison between metasurface simulation using a full-wave solver and the EPIC-Net. While the full-wave solver calculates the field everywhere in three-dimensional space, the EPIC-Net operates on two-dimensional slices adjacent to the metasurface.
    (c) Illustration of how translation equivariance and locality enables seamless generalization from small simulated patches to arbitrarily large metasurfaces.
    (d) Overview of the architecture of the EPIC-Net.
    }\label{fig:overview}
\end{figure*}

Here we show that a neural network architecture tailored to the physics of metasurfaces enables large-scale modeling of two-dimensional free-form metasurfaces under arbitrary illumination while retaining nonlocal interactions and without requiring large-scale simulations as training data.
Integrating the network in a fully differentiable pipeline from metasurface pattern to figure of merit allows us to use standard topology optimization routines to optimize the metasurface for any desired functionality while satisfying fabrication constraints \cite{zhou_minimum_2015, hammond_photonic_2021}. 
By overcoming the bottleneck of computational cost in large-scale inverse design, this approach enables metasurfaces to be designed holistically, rather than as assemblies of locally optimized elements, as illustrated in Fig.~\ref{fig:overview}a. 

We validate our approach along two complementary axes. 
First, we demonstrate accurate inverse design in challenging electromagnetic regimes by realizing high-angle diffraction gratings, where strong phase gradients and nonlocal effects limit the validity of approximate models. 
Second, we demonstrate the scalability of the approach by designing a \SI{1}{mm^2} metahologram more than 1,000 wavelengths across.

\subsection*{Neural network architecture}
The enabling innovation behind this work is the architecture of the neural network surrogate model.
As illustrated in Fig.~\ref{fig:overview}b, the network takes as input the incident field on a plane immediately before the metasurface, as well as an image representing the metasurface pattern.
Using this information, the network predicts the outgoing reflected and transmitted fields on planes adjacent to either side of the metasurface.
We use a complex-valued field representations containing both x and y polarized components, allowing for a fully vectorial description of the metasurface response.
The electromagnetic field at any non-zero distance from the metasurface can be determined using free-space propagation methods.

The architecture of the network is shown in Fig.~\ref{fig:overview}d.
A linear field encoder $\mathcal{E}_\mathbf{E}$ and a nonlinear pattern encoder $\mathcal{E}_\rho$ produce high-dimensional latent descriptions of the incoming field $\mathbf{E}_\mathrm{in}$ and pattern $\rho$.
The latent field is fed to a recurrent unit, which first applies a linear operation $\mathcal{P}$ with learned weights, then applies a linear operation $\mathcal{I} = \mathcal{I}(\rho)$ whose weights are parameterized by the latent pattern, and finally adds the result to a cumulative buffer.
After $N$ iterations of the recurrent unit with shared weights, the cumulative latent field is fed to a linear field decoder $\mathcal{D}_\mathbf{E}$ which produces the outgoing field $\mathbf{E}_\mathrm{out}$.
The resulting expression for the outgoing field is thus
\begin{equation*}
    \mathbf{E}_\mathrm{out} = \mathcal{D}_\mathbf{E}[1 + \mathcal{I}\mathcal{P} + (\mathcal{I}\mathcal{P})^2 + \ldots + (\mathcal{I}\mathcal{P})^N]\mathcal{E}_\mathbf{E} \mathbf{E}_\mathrm{in}.
\end{equation*}
This has a form similar to that of a Dyson series expansion of the electromagnetic Green's operator
\begin{equation*}
    \mathcal{G} = \mathcal{G}_0[1 + \chi\mathcal{G}_0 + (\chi\mathcal{G}_0)^2 + \ldots],
\end{equation*}
where $\mathcal{G}_0$ is the free-space Green's operator and $\chi$ is the material susceptibility.
Based on this analogy, we refer to $\mathcal{P}$ and $\mathcal{I}$ as \textit{propagation} and \textit{interaction} steps, and refer to the recurrent block as a Propagation-Interaction unit (PI-unit).
Note that the map from $\mathbf{E}_\mathrm{in}$ to $\mathbf{E}_\mathrm{out}$ is linear, ensuring correct generalization to linear combinations of the input field.

The architecture is further specialized to the physics of metasurfaces through the implementation of $\mathcal{E}_\mathbf{E}$, $\mathcal{E}_\rho$, $\mathcal{P}$, and $\mathcal{I}$.
We exploit translation equivariance and locality of interactions by implementing $\mathcal{E}_\mathbf{E}$, $\mathcal{E}_\rho$, and $\mathcal{P}$ as convolutions, and $\mathcal{I}$ as a local element-wise multiplication of the latent field and latent pattern.
The network consequently applies the same computation with a finite receptive field (illustrated by yellow circles in Fig.~\ref{fig:overview}b) to calculate the output field at every location on the metasurface.
Crucially, this makes the model generalize naturally from small training patches to entire metasurfaces \cite{bulusu_generalization_2021, khoram_graph_2023, wu_general_2025, xu_deep-learning-enabled_2025, grand_design_2026} as shown in Fig.~\ref{fig:overview}c, allowing it to surpass the size limitations of the full-wave solver it has been trained on.
We furthermore enforce rotation, reflection, and complex phase equivariance using complex group convolutions \cite{cohen_group_2016, lilja_symmetry-informed_2026}, yielding a highly data-efficient model with only 16,028 trainable parameters.
We refer to this architecture as an Equivariant Propagation-Interaction Convolutional Neural Network (EPIC-Net).
A detailed schematic of the architecture is provided in Extended Data Fig.~\ref{fig:architecture}.

\subsection*{Validation on randomly generated samples}
\begin{figure*}[hbt!]
    \centering
    \includegraphics[width=\textwidth]{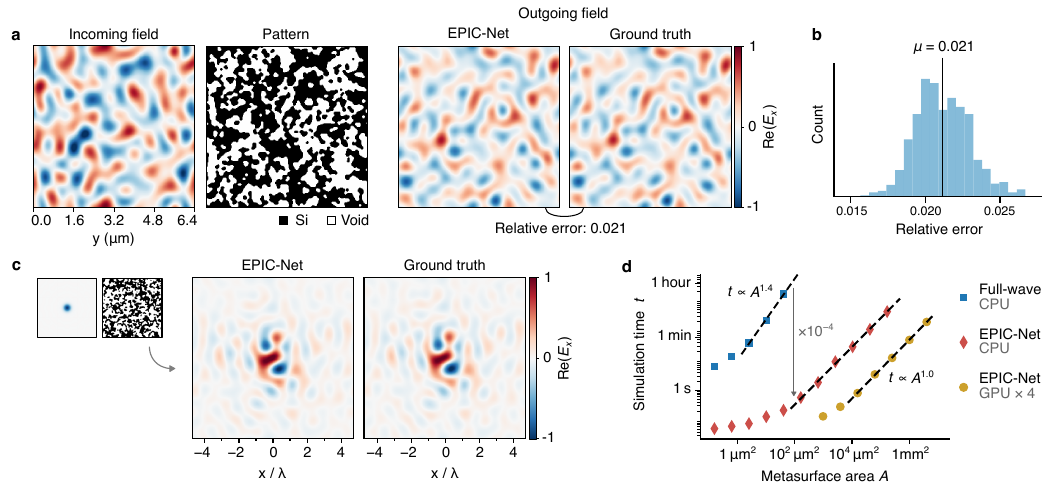}
    \caption{
    Model validation on random metasurface samples. 
    (a) Comparison of the neural network prediction with the ground truth for a sample from the test set. 
    (b) Distribution of errors across the test set, showing that the sample is representative.
    (c) Comparison between neural network prediction and ground truth when the input field is a tightly focused Gaussian beam, demonstrating nonlocal interactions in the metasurface. 
    (d) Simulation time as a function of metasurface area for a full-wave solver and the EPIC-Net.
    }
    \label{fig:validation}
\end{figure*}

The EPIC-Net is trained on approximately 10,000 simulated samples of dielectric metasurfaces with randomly generated patterns and incoming fields, with the corresponding outgoing fields generated by a full-wave finite-element frequency-domain solver which we consider as the ground truth.
The simulations assume a free-space wavelength of $980\,$nm in a surrounding medium with refractive index 1.45, resulting in an effective wavelength of $\lambda = 675.9\,$nm.
We use $\SI{3.2}{\micro\meter} \times \SI{3.2}{\micro\meter}$ patches for training, while validating and testing on larger $\SI{6.4}{\micro\meter} \times \SI{6.4}{\micro\meter}$ patches to verify generalization to larger areas.
Figure~\ref{fig:validation}a shows one such test sample and the corresponding network prediction, demonstrating good agreement with the ground-truth solver.
The relative error of this sample is 2.1\%, which is representative of the distribution of errors across the test set, shown in Fig.~\ref{fig:validation}b.
The error metric is the mean-squared error of the normalized E-field, which can be interpreted as the intensity of the error relative to the total intensity of the outgoing field.

The EPIC-Net is able to capture interactions in the metasurface that go beyond local response and slowly varying field approximations.
To illustrate this, we show in Fig.~\ref{fig:validation}c the outgoing fields predicted by the EPIC-Net and the ground-truth solver for a randomly generated metasurface pattern when the incoming field is a tightly focused Gaussian beam.
The outgoing field exhibits significant variations more than four wavelengths from the center of the incoming beam, in clear violation of local response. 
Nevertheless, the EPIC-Net reproduces the entire outgoing field with high accuracy.

The benefit of the EPIC-Net compared to the full-wave solver is computational efficiency, as shown in Fig.~\ref{fig:validation}d.
The EPIC-Net is more than four orders of magnitude faster than the reference solver for areas larger than \SI{100}{\micro\meter\squared} when running on the same 64-core CPU.
The EPIC-Net inference time also scales proportionally with the metasurface area, whereas the reference solver scales superlinearly, which is typical for frequency-domain solvers.
Furthermore, being a convolutional neural network, the EPIC-Net is ideally suited to run on hardware accelerators for deep learning.
Parallelizing inference across four A100 GPUs enables millimeter-scale metasurfaces to be simulated in minutes, making iterative topology optimization at this scale practical.

\subsection*{Inverse design of metagratings}
\begin{figure}[hbt!]
    \centering
    \includegraphics[width=0.5\textwidth]{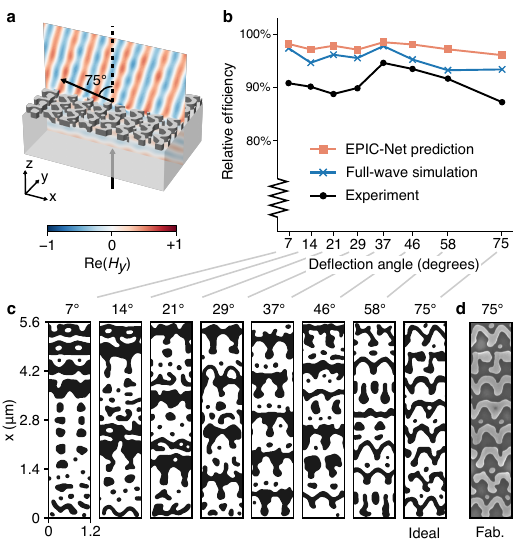}
    \caption{
    Inverse-designed metagratings. 
    (a) Ground-truth simulation of a grating inverse-designed using the EPIC-Net. 
    (b) Relative efficiency as a function of deflection angle according to the neural network prediction, full-wave solver, and experimental measurement. 
    (c) Inverse-designed patterns of the unit cells of the fabricated gratings.
    (d) SEM image of the fabricated grating corresponding to a deflection angle of 75 degrees.}
    \label{fig:gratings}
\end{figure}

To further validate the accuracy of the EPIC-Net, we apply it to the inverse design of metagratings, a well-studied system in metasurface topology optimization \cite{sell_large-angle_2017}.
Metagratings are periodic metasurfaces that deflect light at an angle, as illustrated in Fig.~\ref{fig:gratings}a.
We fix the lattice constant of the gratings in the x direction at slightly more than 8 free-space wavelengths, allowing for eight diffraction orders with angles ranging from 7 to 75 degrees.
We then run a standard topology optimization procedure (illustrated in Extended Data Fig.~\ref{fig:topopt}) using the EPIC-Net as a surrogate for full-wave simulations to generate eight gratings, each maximizing the fraction of outgoing light for one of the angles.
Eight candidate designs per angle were generated, and the best-performing designs according to the full-wave solver were selected for fabrication.

Figure~\ref{fig:gratings}b shows the network-predicted, full-wave simulated, and experimentally measured relative efficiency (fraction of transmitted power in the desired mode) of the fabricated metagratings as a function of the deflection angle.
The EPIC-Net-predicted efficiency overestimates the full-wave simulated efficiency by a few percent, likely because the topology optimization exploits inaccuracies of the model.
Nevertheless, our neural-network-accelerated method is able to find gratings with ground-truth efficiencies around 95 percent for deflection angles up to at least 75 degrees, on par with more computationally expensive full-wave topology optimization \cite{sell_large-angle_2017}.
The experimental efficiencies are around 5 percent lower than the full-wave solver predictions.
This is likely due to fabrication imperfections causing differences between the ideal simulated patterns in Fig.~\ref{fig:gratings}c and the fabricated patterns, of which a scanning electron microscopy (SEM) image is shown in Fig.~\ref{fig:gratings}d.
Absolute efficiencies and SEM images of the other gratings are provided in Extended Data Fig.~\ref{fig:grating_extra}.

\subsection*{Inverse design of a large-area metahologram}
\begin{figure*}[hbt!]
    \centering
    \includegraphics[width=\textwidth]{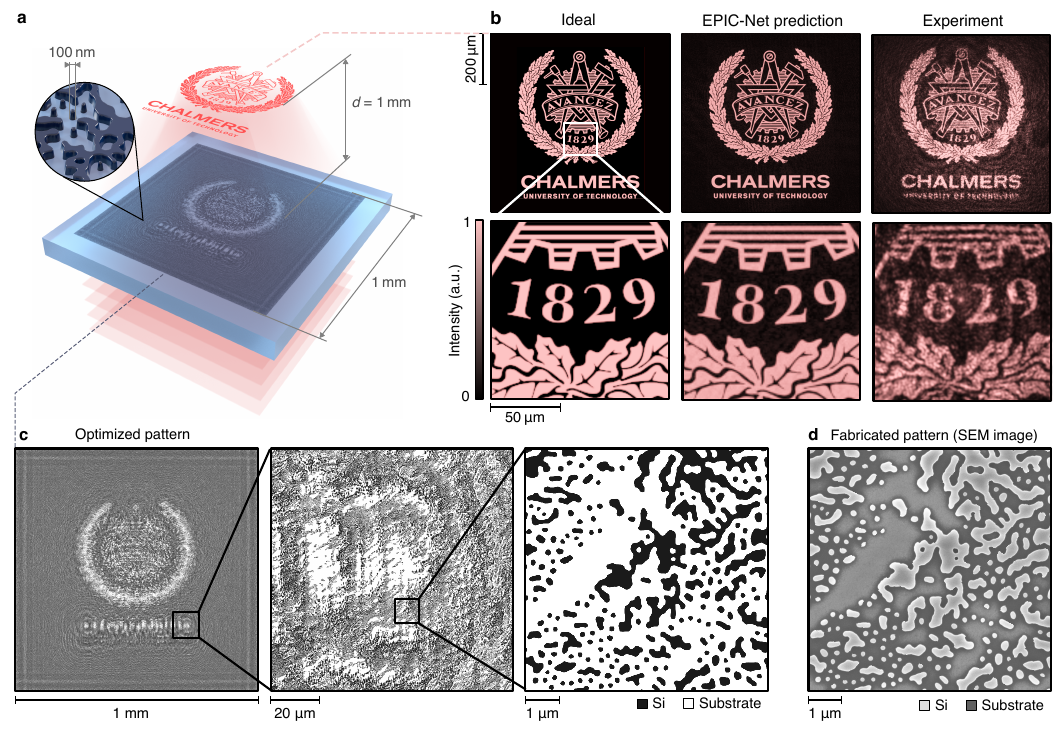}
    \caption{
    Inverse-designed macroscopic free-form metahologram. 
    (a) Artistic illustration of the hologram geometry.
    (b) Comparison of the ideal, network-predicted, and measured intensity distributions in the image plane.
    (c) The inverse-designed hologram pattern shown at three different magnification levels.    
    (d) SEM image of the fabricated pattern.}
    \label{fig:hologram}
\end{figure*}

Finally, we demonstrate the scalability of our approach by designing a fully free-form metasurface hologram 
with an area of \SI{1}{\milli\meter\squared}, corresponding to more than $1000\lambda \times 1000\lambda$ and comprising 1.6 billion degrees of freedom (pixels in the pattern bitmap).
This exceeds the area of the largest previously reported two-dimensional non-symmetric free-form metasurface that we were able to find ($80\lambda\times80\lambda$ utilizing a locally periodic approximation \cite{lin_topology_2019}) by more than two orders of magnitude.
The hologram, illustrated in Fig.~\ref{fig:hologram}a, is designed to generate a predefined intensity distribution in an image plane 1\,mm in front of the metasurface.
The metasurface pattern is generated in a single optimization loop where the entire metasurface is simulated using the EPIC-Net at each step, resulting in a seamless free-form pattern bridging the nanoscale and the macroscale.
The area of the hologram is ten times larger than the combined area of the samples in the training dataset, making the computational cost of training the EPIC-Net negligible in comparison to what it would cost to design the hologram using the full-wave solver.

The ideal, network-predicted, and experimentally measured intensity distributions in the image plane are shown in Fig.~\ref{fig:hologram}b.
The EPIC-Net prediction is nearly indistinguishable from the ideal image, illustrating that the richness of responses attainable through fully free-form inverse design is sufficient to generate holograms of excellent quality with no speckle, twin images, or zeroth-order beam.
We validate the hologram through experiment since it is too large to be simulated using the full-wave solver.
The experimentally measured image clearly resembles the ideal image and captures most of its finer details, albeit with some noise due to a combination of neural network, fabrication, and measurement errors.
Detailed images of various parts of the hologram pattern and holographic image, as well as measured intensity distributions at other distances $d$ from the hologram, are provided in Extended Data Fig.~\ref{fig:hologram_extra}.

\section*{Discussion}
In summary, we have presented a neural network architecture tailored to the physics of metasurfaces that enables free-form topology optimization at unprecedented scales.
Numerical and experimental validation confirms that the network is able to capture the intricate electromagnetic interactions required to realize high-performance metagratings, while being scalable to macroscopic metasurfaces with billions of free parameters.

The method presented here has potential applications beyond large-scale metasurface inverse design.
For smaller devices, the computational efficiency of the EPIC-Net makes it possible to run several orders of magnitude more simulations at the same cost. 
We envision that this will be useful for fabrication-robust inverse design \cite{elbek_tailoring_2026}, to co-optimize multiple cascaded metasurfaces \cite{mei_cascaded_2023}, in studies of disordered metasurfaces \cite{hu_disordered_2023}, and to scale up the training of optical neural networks \cite{zhou_optical_2024}.
Our approach may also inform inverse-design strategies in other wave-based systems, including acoustics, microwave engineering, and integrated photonics.

A current limitation is that the receptive radius of the EPIC-Net cannot be made arbitrarily large, preventing accurate modeling of strongly resonant or waveguide-like connected structures that support extremely long-range interactions.
The 2.1\% error reported here was achieved using 64 iterations of the recurrent unit, resulting in a maximum interaction length of \SI{12.8}{\micro\meter}.
Increasing the number of iterations further is currently not possible due to exponentially growing activations in the recurrent unit making predictions unstable.
We believe this effect could be mitigated in future work by applying constraints on the recurrent unit that enforce convergence, potentially in combination with active learning \cite{singh_active_2025} for improved sampling of metasurfaces with long-range interactions.

Another limitation is that the EPIC-Net has so far been trained on a single wavelength, metasurface thickness, and material.
Extending the framework to multiple wavelengths is an important next step to unlock free-form large-scale metasurfaces for achromatic lensing \cite{wang_broadband_2017}, full-color holography \cite{gopakumar_full-color_2024}, and augmented reality \cite{lee_metasurface_2018}.
Making the model applicable to a wider range of fabrication parameters would also be beneficial to facilitate reuse.
Combined with the ability to model any input illumination and any metasurface pattern, this would make the EPIC-Net highly general, potentially allowing it to take the role of a foundation model for metasurfaces.

Looking ahead, we expect that this work will serve as the basis for a new generation of system-scale metadevices that utilize the full design freedom of planar fabrication to push the boundaries of wavefront manipulation.

\section*{Methods}
\subsection*{Metasurface parameters}
All metasurfaces were made of a patterned 242.2\,nm-thick layer of amorphous silicon with a refractive index $3.923+0.008i$, on top of a fused silica substrate with a refractive index of 1.45.
These parameters were determined by polarimetric measurements of the substrate that was later used in the fabrication of the metasurfaces.
The same parameters were used in the full-wave simulations to generate training data and validate the designed metasurfaces.
For all results except the hologram, the metasurface was coated with a protective polymer layer with a refractive index of 1.45 to enable the use of an oil-immersion objective for experimental characterization.
A separate dataset with air (refractive index 1.00) above the metasurface was used for the hologram.
The EPIC-Net performance on the dataset with air is similar to the performance on the dataset with polymer.

\subsection*{Training data generation}
The training datasets consisted of approximately \num{10000} randomly generated pairs of metasurface patterns and incoming fields sampled on a grid with a pixel size of \SI{25}{nm}, along with corresponding outgoing fields generated by the full-wave solver.
Random metasurface patterns were generated by applying a smoothing filter and thresholding to white noise images, then applying morphological opening and closing operations in random order to approximately enforce minimum feature-size constraints.
Incoming fields were generated by sampling random superpositions of plane waves incoming from all directions.
The amplitudes were scaled so that each plane wave and polarization carried the same power on average, and amplitudes of evanescent waves ($k_x^2 + k_y^2 > k^2$) were set to zero.

\subsection*{Full-wave simulations}
A direct finite-element frequency-domain solver implemented in COMSOL Multiphysics was used as the full-wave reference solver.
Periodic boundary conditions were used in the x and y directions and perfectly matched layers were used above and below the metasurface.
To generate the fields in the training data, the computed electromagnetic field was interpolated and sampled on a grid with the same resolution as the pattern and incoming field.
Because the outgoing field was sampled immediately before and after the metasurface (on planes in contact with the patterned region), it contains evanescent components that decay quickly after free-space propagation. 
Since evanescent waves are not relevant for the problems studied here, they were removed by applying a frequency-domain filter post-sampling.

The full-wave solver was chosen based on its robustness and precision for non-rectilinear free-form geometries.
We acknowledge that more computationally efficient full-wave solvers are available and would make the comparative speedup shown in Fig.~\ref{fig:validation}d smaller.
However, the EPIC-Net retains an advantage of two orders of magnitude even when compared to state-of-the-art GPU-accelerated finite-element time-domain methods \cite{hughes_perspective_2021}.

\subsection*{Neural network architecture}
A detailed schematic of the EPIC-Net architecture is shown in Extended Data Fig.~\ref{fig:architecture}.
The field encoder and decoder consist of resampling operations and pointwise convolutions.
According to the Nyquist-Shannon sampling theorem, the resampling incurs no loss of information since both the incoming field and the target outgoing fields are bandlimited after evanescent components have been removed.
The pattern encoder consists of a small CNN with ReLU activation and downsampling to make the latent pattern match the resolution of the latent field.
We found downsampling by a factor of four to be optimal for our datasets.
The propagation step is implemented as a convolution with kernel size $5\times5$.
$D_4$-equivariance is enforced using group convolutions implemented in escnn \cite{cesa_program_2022}.
Note that the interaction step respects $D_4$ equivariance as long as the latent pattern and latent field both transform according to the regular representation.
All linear operations on the field use complex numbers to ensure that the map from the incoming field to the outgoing field respects complex phase equivariance.
All convolutions use periodic padding to make the network match the periodic boundary conditions of the training data.
A spatial filter removing evanescent components from the output field was included as a final layer to match the training data.

\subsection*{Neural network training}
The network was trained for 25 epochs using the Adam optimizer.
To facilitate convergence, we started with a large learning rate and only four iterations of the PI-unit, then doubled the number of iterations and reduced the learning rate by a fixed factor every five epochs.
A small amount of white noise was added to the pattern bitmaps during training.
The initial learning rate, learning rate decay factor, weight decay, and noise amplitude hyperparameters were optimized using Optuna \cite{akiba_optuna_2019}.
Translation data augmentation was applied since downsampling of the metasurface pattern breaks equivariance to translations that are not an integer multiple of four pixels \cite{lilja_symmetry-informed_2026}.

\subsection*{Topology optimization algorithm}
The metasurfaces were designed by integrating the EPIC-Net into a standard topology optimization routine \cite{jensen_topology_2011, hammond_photonic_2021} illustrated in Extended Data Fig.~\ref{fig:topopt}a.
The metasurface design is represented by a density field $\rho \in [0,1]$.
At each topology optimization step, the density is first convolved with a conic filter $\tilde{\rho} = w*\rho$, then binarized by a tanh-projection 
\begin{equation*}
\bar{\rho} = \frac{\tanh\left(\beta\eta\right)+\tanh\left(\beta\left(\tilde{\rho}-\eta\right)\right)}{\tanh\left(\beta\eta\right)+\tanh\left(\beta\left(1-\eta\right)\right)}    
\end{equation*}
with $\eta=1/2$.
The binarized pattern $\bar{\rho}$ is given as an input to a pre-trained EPIC-Net along with the fixed input field $\mathbf{E}_\mathrm{in}$ to yield the outgoing field $\mathbf{E}_\mathrm{out}$, from which an objective value is calculated.
The objective value calculation may include a free-space propagation step and pattern-dependent terms.
The gradient of the objective is backpropagated to the design field $\rho$ using automatic differentiation and used by an optimizer to update the design.
The cycle is repeated to gradually improve the design while slowly increasing $\beta$ to enforce binarization.

\subsection*{Topology optimization parameters}
We used the Adam optimizer \cite{kingma_adam_2015} for topology optimization with 600 optimization steps for the gratings and 150 steps for the hologram.
Constraints were enforced approximately by adding a large penalty to the objective value for constraint violations.
Additional penalty terms were added to enforce a minimum linewidth and minimum spacing of 100\,nm \cite{hammond_photonic_2021}.
The figure of merit for the gratings was defined as the overlap of the outgoing field with an ideal output field with all power in the desired diffraction order.
The figure of merit for the hologram was defined as the cross-correlation between the target image and the z-component of the time-averaged Poynting vector in the image plane.
The field was propagated from the metasurface to the image plane using the angular spectrum method.
We use an optimization schedule in which $\beta$ increases gradually during the second third of the optimization (steps 200 to 400 for the gratings), and the feature size penalty increases gradually during the last third (steps 400 to 600).
The evolution of a sample grating is shown in Extended Data Fig.~\ref{fig:topopt}b.

\subsection*{Vectorization of raster patterns}
Before fabrication and full-wave simulation, metasurface patterns were converted from a raster representation to a vectorized representation (list of polygons) by applying a Gaussian smoothing filter followed by the marching squares algorithm.
Extended Data Fig.~\ref{fig:topopt}c illustrates the conversion and shows how the vectorized pattern corresponds to the simulated design.
The same procedure was used to generate the simulated geometries in the training dataset, meaning that the vectorization operation is taken into account by the trained EPIC-Net. Therefore, no additional error is introduced by the conversion.

\subsection*{Handling of grayscale densities}
During the early stages of topology optimization, the pattern density field $\rho$ typically contains grayscale values, i.e., densities between those corresponding to void and silicon.
Unlike in conventional topology optimization, we make no explicit choice about how such intermediate densities should be interpreted.
Only binarized metasurfaces were simulated by the full-wave solver, and no grayscale patterns were present in the training dataset.
The response to intermediate densities is therefore determined only by how the neural network interpolates between its training samples.
We find that our strongly physics-constrained architecture makes this interpolation sufficiently well-behaved for topology optimization to converge.
Since the patterns are nearly perfectly binarized at the end of the topology optimization, the response of the final optimized metasurface is always well-defined.

\subsection*{Memory management and parallelization}
During topology optimization of the hologram, we divided the metasurface into 400 patches of $\SI{50}{\micro\meter}\times\SI{50}{\micro\meter}$, padded so that they partially overlap, and calculated the output field for each patch independently, with the calculations parallelized across four GPUs.
This partitioning does not change the final predicted output field and therefore incurs no loss of accuracy as long as the overlap exceeds the receptive-field radius of the network.
A gradient checkpointing scheme was employed to avoid storing intermediate activations of the neural network, which are needed for backpropagation.
This scheme makes the memory requirement of network evaluation during topology optimization independent of metasurface size, at the cost of requiring an additional forward pass for each optimization step.
The only operation that must be performed at the full-metasurface level is evaluation and backpropagation through the objective function, which in our case could be done on a single GPU.

\subsection*{Metasurface fabrication}
The metasurfaces were fabricated using conventional electron-beam lithography (EBL). 
The amorphous silicon layer was deposited onto the fused silica substrate via low-pressure chemical vapor deposition (LPCVD). 
A 200~nm layer of AR-P6200.13 served as the electron-beam resist. 
Following pattern exposure and development, a metal mask was deposited using an electron-beam evaporator. 
After a lift-off process, the pattern was transferred to the silicon film through inductively coupled plasma (ICP) dry etching. 
The metal mask was removed via wet etching. 
Finally, the grating metasurfaces were coated with a protective polymer layer.

\subsection*{Grating measurements}
The experimental setup for the metagrating measurements is illustrated in Extended Data Fig.~\ref{fig:experimental_setup}a.
The collimated and expanded beam from a 976~nm butterfly fibre-coupled laser was used to illuminate the samples. 
The transmitted light was collected by a 100x oil-immersion objective (NA = 1.49). 
An aperture was placed in the conjugate plane of the grating to block light going around the sides of the metasurface.
Note that this does not block the transmitted intensity of the zeroth-order beam through the grating.
Diffraction patterns were obtained by imaging the back focal plane of the objective.

\subsection*{Metagrating data analysis}
From the diffraction patterns, the relative power in each order was calculated by integrating the measured intensity over a circular region around the expected location of each diffraction peak, as illustrated in Extended Data Fig.~\ref{fig:grating_extra}a.
We additionally performed a measurement for an empty substrate to measure the total incoming power.
Extended Data Fig.~\ref{fig:grating_extra}a shows the measured intensity distribution, as well as the calculated relative efficiency (power in the desired mode divided by power in all transmitted modes) and absolute efficiency (power in the desired mode divided by incoming power) for each grating.
The optimized grating geometries are shown next to SEM images of the fabricated gratings in Extended Data Fig.~\ref{fig:grating_extra}c and d.

\subsection*{Hologram measurements}
The experimental setup for the hologram measurements is illustrated in Extended Data Fig.~\ref{fig:experimental_setup}b.
The hologram was illuminated by a beam similar to that used for the grating measurements. 
An objective was positioned such that the holographic image plane (appearing at a distance of \SI{1}{mm} from the metasurface) coincided with its object plane. 
Images were recorded by a detector placed in a conjugate plane of the holographic image formed by an imaging lens.
Objectives with 20x or 60x (NA = 0.5 and 0.7, respectively) were used to obtain images of with different magnification levels. 

\section*{Data availability}
The datasets generated during and analysed during the current study are available from the corresponding author on reasonable request.

\section*{Code availability}
Data were processed and analysed as described in the Methods and main text. The code is available from the corresponding author on reasonable request.

%\bibliographystyle{naturemag}
%\bibliography{references}

\begin{thebibliography}{10}
\expandafter\ifx\csname url\endcsname\relax
  \def\url#1{\texttt{#1}}\fi
\expandafter\ifx\csname urlprefix\endcsname\relax\def\urlprefix{URL }\fi
\providecommand{\bibinfo}[2]{#2}
\providecommand{\eprint}[2][]{\url{#2}}

\bibitem{yu_flat_2014}
\bibinfo{author}{Yu, N.} \& \bibinfo{author}{Capasso, F.}
\newblock \bibinfo{title}{Flat optics with designer metasurfaces}.
\newblock \emph{\bibinfo{journal}{Nat. Mater.}} \textbf{\bibinfo{volume}{13}}, \bibinfo{pages}{139--150} (\bibinfo{year}{2014}).

\bibitem{khorasaninejad_metalenses_2017}
\bibinfo{author}{Khorasaninejad, M.} \& \bibinfo{author}{Capasso, F.}
\newblock \bibinfo{title}{Metalenses: {{Versatile}} multifunctional photonic components}.
\newblock \emph{\bibinfo{journal}{Science}} \textbf{\bibinfo{volume}{358}}, \bibinfo{pages}{eaam8100} (\bibinfo{year}{2017}).

\bibitem{zeng_performance_2025}
\bibinfo{author}{Zeng, Y.}, \bibinfo{author}{Zhong, H.}, \bibinfo{author}{Long, Z.}, \bibinfo{author}{Cao, H.} \& \bibinfo{author}{Jin, X.}
\newblock \bibinfo{title}{From performance to structure: A comprehensive survey of advanced metasurface design for next-generation imaging}.
\newblock \emph{\bibinfo{journal}{npj Nanophoton.}} \textbf{\bibinfo{volume}{2}}, \bibinfo{pages}{39} (\bibinfo{year}{2025}).

\bibitem{jensen_topology_2011}
\bibinfo{author}{Jensen, J.~S.} \& \bibinfo{author}{Sigmund, O.}
\newblock \bibinfo{title}{Topology optimization for nano-photonics}.
\newblock \emph{\bibinfo{journal}{Laser Photonics Rev.}} \textbf{\bibinfo{volume}{5}}, \bibinfo{pages}{308--321} (\bibinfo{year}{2011}).

\bibitem{kuznetsov_roadmap_2024}
\bibinfo{author}{Kuznetsov, A.~I.} \emph{et~al.}
\newblock \bibinfo{title}{Roadmap for {{Optical Metasurfaces}}}.
\newblock \emph{\bibinfo{journal}{ACS Photonics}} \textbf{\bibinfo{volume}{11}}, \bibinfo{pages}{816--865} (\bibinfo{year}{2024}).

\bibitem{schul_inverse_2026}
\bibinfo{author}{Schul, L.}, \bibinfo{author}{Mason, S.}, \bibinfo{author}{Eun, S.}, \bibinfo{author}{Ahn, G.~H.} \& \bibinfo{author}{Vučković, J.}
\newblock \bibinfo{title}{Inverse design for scalable photonic systems}.
\newblock \emph{\bibinfo{journal}{Nat. Rev. Mater.}} \bibinfo{pages}{1--11} (\bibinfo{year}{2026}).

\bibitem{sell_large-angle_2017}
\bibinfo{author}{Sell, D.}, \bibinfo{author}{Yang, J.}, \bibinfo{author}{Doshay, S.}, \bibinfo{author}{Yang, R.} \& \bibinfo{author}{Fan, J.~A.}
\newblock \bibinfo{title}{Large-{{Angle}}, {{Multifunctional Metagratings Based}} on {{Freeform Multimode Geometries}}}.
\newblock \emph{\bibinfo{journal}{Nano Lett.}} \textbf{\bibinfo{volume}{17}}, \bibinfo{pages}{3752--3757} (\bibinfo{year}{2017}).

\bibitem{phan_high-efficiency_2019}
\bibinfo{author}{Phan, T.} \emph{et~al.}
\newblock \bibinfo{title}{High-efficiency, large-area, topology-optimized metasurfaces}.
\newblock \emph{\bibinfo{journal}{Light Sci. Appl.}} \textbf{\bibinfo{volume}{8}}, \bibinfo{pages}{48} (\bibinfo{year}{2019}).

\bibitem{liu_topology_2025}
\bibinfo{author}{Liu, Y.-T.} \emph{et~al.}
\newblock \bibinfo{title}{Topology {{Optimization Enables Freeform Matrix Fourier Optics}}}.
\newblock \emph{\bibinfo{journal}{Adv. Opt. Mater.}} \textbf{\bibinfo{volume}{13}}, \bibinfo{pages}{2500872} (\bibinfo{year}{2025}).

\bibitem{sun_scalable_2025}
\bibinfo{author}{Sun, M.} \emph{et~al.}
\newblock \bibinfo{title}{Scalable {{Freeform Optimization}} of {{Wide-Aperture 3D Metalenses}} by {{Zoned Discrete Axisymmetry}}}.
\newblock \emph{\bibinfo{journal}{ACS Photonics}} \textbf{\bibinfo{volume}{12}}, \bibinfo{pages}{3163--3171} (\bibinfo{year}{2025}).

\bibitem{choi_realization_2024}
\bibinfo{author}{Choi, M.} \emph{et~al.}
\newblock \bibinfo{title}{Realization of high-performance optical metasurfaces over a large area: A review from a design perspective}.
\newblock \emph{\bibinfo{journal}{npj Nanophoton.}} \textbf{\bibinfo{volume}{1}}, \bibinfo{pages}{31} (\bibinfo{year}{2024}).

\bibitem{kang_large-scale_2024}
\bibinfo{author}{Kang, C.} \emph{et~al.}
\newblock \bibinfo{title}{Large-scale photonic inverse design: Computational challenges and breakthroughs}.
\newblock \emph{\bibinfo{journal}{Nanophotonics}} \textbf{\bibinfo{volume}{13}}, \bibinfo{pages}{3765--3792} (\bibinfo{year}{2024}).

\bibitem{hughes_perspective_2021}
\bibinfo{author}{Hughes, T.~W.}, \bibinfo{author}{Minkov, M.}, \bibinfo{author}{Liu, V.}, \bibinfo{author}{Yu, Z.} \& \bibinfo{author}{Fan, S.}
\newblock \bibinfo{title}{A perspective on the pathway toward full wave simulation of large area metalenses}.
\newblock \emph{\bibinfo{journal}{Appl. Phys. Lett.}} \textbf{\bibinfo{volume}{119}}, \bibinfo{pages}{150502} (\bibinfo{year}{2021}).

\bibitem{lin_topology_2019}
\bibinfo{author}{Lin, Z.}, \bibinfo{author}{Liu, V.}, \bibinfo{author}{Pestourie, R.} \& \bibinfo{author}{Johnson, S.~G.}
\newblock \bibinfo{title}{Topology optimization of freeform large-area metasurfaces}.
\newblock \emph{\bibinfo{journal}{Opt. Express}} \textbf{\bibinfo{volume}{27}}, \bibinfo{pages}{15765--15775} (\bibinfo{year}{2019}).

\bibitem{lin_overlapping_2019}
\bibinfo{author}{Lin, Z.} \& \bibinfo{author}{Johnson, S.~G.}
\newblock \bibinfo{title}{Overlapping domains for topology optimization of large-area metasurfaces}.
\newblock \emph{\bibinfo{journal}{Opt. Express}} \textbf{\bibinfo{volume}{27}}, \bibinfo{pages}{32445--32453} (\bibinfo{year}{2019}).

\bibitem{nadell_deep_2019}
\bibinfo{author}{Nadell, C.~C.}, \bibinfo{author}{Huang, B.}, \bibinfo{author}{Malof, J.~M.} \& \bibinfo{author}{Padilla, W.~J.}
\newblock \bibinfo{title}{Deep learning for accelerated all-dielectric metasurface design}.
\newblock \emph{\bibinfo{journal}{Opt. Express}} \textbf{\bibinfo{volume}{27}}, \bibinfo{pages}{27523--27535} (\bibinfo{year}{2019}).

\bibitem{an_deep_2019}
\bibinfo{author}{An, S.} \emph{et~al.}
\newblock \bibinfo{title}{A {{Deep Learning Approach}} for {{Objective-Driven All-Dielectric Metasurface Design}}}.
\newblock \emph{\bibinfo{journal}{ACS Photonics}} \textbf{\bibinfo{volume}{6}}, \bibinfo{pages}{3196--3207} (\bibinfo{year}{2019}).

\bibitem{hegde_deep_2020}
\bibinfo{author}{Hegde, R.~S.}
\newblock \bibinfo{title}{Deep learning: A new tool for photonic nanostructure design}.
\newblock \emph{\bibinfo{journal}{Nanoscale Adv.}} \textbf{\bibinfo{volume}{2}}, \bibinfo{pages}{1007--1023} (\bibinfo{year}{2020}).

\bibitem{wiecha_deep_2021}
\bibinfo{author}{Wiecha, P.~R.}, \bibinfo{author}{Arbouet, A.}, \bibinfo{author}{Girard, C.} \& \bibinfo{author}{Muskens, O.~L.}
\newblock \bibinfo{title}{Deep learning in nano-photonics: Inverse design and beyond}.
\newblock \emph{\bibinfo{journal}{Photonic Res.}} \textbf{\bibinfo{volume}{9}}, \bibinfo{pages}{B182} (\bibinfo{year}{2021}).

\bibitem{so_deep_2020}
\bibinfo{author}{So, S.}, \bibinfo{author}{Badloe, T.}, \bibinfo{author}{Noh, J.}, \bibinfo{author}{{Bravo-Abad}, J.} \& \bibinfo{author}{Rho, J.}
\newblock \bibinfo{title}{Deep learning enabled inverse design in nanophotonics}.
\newblock \emph{\bibinfo{journal}{Nanophotonics}} \textbf{\bibinfo{volume}{9}}, \bibinfo{pages}{1041--1057} (\bibinfo{year}{2020}).

\bibitem{khaireh-walieh_newcomers_2023}
\bibinfo{author}{{Khaireh-Walieh}, A.} \emph{et~al.}
\newblock \bibinfo{title}{A newcomer’s guide to deep learning for inverse design in nano-photonics}.
\newblock \emph{\bibinfo{journal}{Nanophotonics}} \textbf{\bibinfo{volume}{12}}, \bibinfo{pages}{4387--4414} (\bibinfo{year}{2023}).

\bibitem{augenstein_neural_2023}
\bibinfo{author}{Augenstein, Y.}, \bibinfo{author}{Repa\"an, T.} \& \bibinfo{author}{Rockstuhl, C.}
\newblock \bibinfo{title}{Neural operator-based surrogate solver for free-form electromagnetic inverse design}.
\newblock \emph{\bibinfo{journal}{ACS Photonics}} \textbf{\bibinfo{volume}{10}}, \bibinfo{pages}{1547--1557} (\bibinfo{year}{2023}).

\bibitem{gahlmann_deep_2022}
\bibinfo{author}{Gahlmann, T.} \& \bibinfo{author}{Tassin, P.}
\newblock \bibinfo{title}{Deep neural networks for the prediction of the optical properties and the free-form inverse design of metamaterials}.
\newblock \emph{\bibinfo{journal}{Phys. Rev. B}} \textbf{\bibinfo{volume}{106}}, \bibinfo{pages}{085408} (\bibinfo{year}{2022}).

\bibitem{furat_physics-informed_2025}
\bibinfo{author}{Furat, O.}, \bibinfo{author}{Gogineni, V.~C.}, \bibinfo{author}{Bindslev, H.} \& \bibinfo{author}{Nadimi, E.~S.}
\newblock \bibinfo{title}{Physics-informed {{Neural Operators}} for {{Predicting 3D Electromagnetic Fields Transformed}} by {{Metasurfaces}}}.
\newblock \bibinfo{howpublished}{Preprint at https://doi.org/10.48550/arXiv.2512.15694} (\bibinfo{year}{2025}).

\bibitem{majorel_deep_2022}
\bibinfo{author}{Majorel, C.}, \bibinfo{author}{Girard, C.}, \bibinfo{author}{Arbouet, A.}, \bibinfo{author}{Muskens, O.~L.} \& \bibinfo{author}{Wiecha, P.~R.}
\newblock \bibinfo{title}{Deep {{Learning Enabled Strategies}} for {{Modeling}} of {{Complex Aperiodic Plasmonic Metasurfaces}} of {{Arbitrary Size}}}.
\newblock \emph{\bibinfo{journal}{ACS Photonics}} \textbf{\bibinfo{volume}{9}}, \bibinfo{pages}{575--585} (\bibinfo{year}{2022}).

\bibitem{an_deep_2022}
\bibinfo{author}{An, S.} \emph{et~al.}
\newblock \bibinfo{title}{Deep {{Convolutional Neural Networks}} to {{Predict Mutual Coupling Effects}} in {{Metasurfaces}}}.
\newblock \emph{\bibinfo{journal}{Adv. Opt. Mater.}} \textbf{\bibinfo{volume}{10}}, \bibinfo{pages}{2102113} (\bibinfo{year}{2022}).

\bibitem{chen_artificial_2022}
\bibinfo{author}{Chen, M.~K.}, \bibinfo{author}{Liu, X.}, \bibinfo{author}{Sun, Y.} \& \bibinfo{author}{Tsai, D.~P.}
\newblock \bibinfo{title}{Artificial {{Intelligence}} in {{Meta-optics}}}.
\newblock \emph{\bibinfo{journal}{Chem. Rev.}} \textbf{\bibinfo{volume}{122}}, \bibinfo{pages}{15356--15413} (\bibinfo{year}{2022}).

\bibitem{khoram_graph_2023}
\bibinfo{author}{Khoram, E.}, \bibinfo{author}{Wu, Z.}, \bibinfo{author}{Qu, Y.}, \bibinfo{author}{Zhou, M.} \& \bibinfo{author}{Yu, Z.}
\newblock \bibinfo{title}{Graph {{Neural Networks}} for {{Metasurface Modeling}}}.
\newblock \emph{\bibinfo{journal}{ACS Photonics}} \textbf{\bibinfo{volume}{10}}, \bibinfo{pages}{892--899} (\bibinfo{year}{2023}).

\bibitem{wu_general_2025}
\bibinfo{author}{Wu, O.}, \bibinfo{author}{Qian, C.}, \bibinfo{author}{Fan, Z.}, \bibinfo{author}{Zhu, X.} \& \bibinfo{author}{Chen, H.}
\newblock \bibinfo{title}{General {{Characterization}} of {{Intelligent Metasurfaces}} with {{Graph Coupling Network}}}.
\newblock \emph{\bibinfo{journal}{Laser Photonics Rev.}} \textbf{\bibinfo{volume}{19}}, \bibinfo{pages}{2400979} (\bibinfo{year}{2025}).

\bibitem{xu_deep-learning-enabled_2025}
\bibinfo{author}{Xu, B.} \emph{et~al.}
\newblock \bibinfo{title}{Deep-learning-enabled inverse design of large-scale metasurfaces with full-wave accuracy}.
\newblock \emph{\bibinfo{journal}{Laser Photonics Rev.}} \textbf{\bibinfo{volume}{20}}, \bibinfo{pages}{e03115} (\bibinfo{year}{2025}).

\bibitem{grand_design_2026}
\bibinfo{author}{Grand, M.~L.}
\newblock \bibinfo{title}{Design and {{Optimization}} of {{Metasurfaces}} for {{Silicon Photonics}}: {{PhD Thesis}}}.
\newblock \bibinfo{howpublished}{Available at https://doi.org/10.48550/arxiv.2607.21091} (\bibinfo{year}{2026}).

\bibitem{lim_maxwellnet_2022}
\bibinfo{author}{Lim, J.} \& \bibinfo{author}{Psaltis, D.}
\newblock \bibinfo{title}{Maxwellnet: Physics-driven deep neural network training based on maxwell’s equations}.
\newblock \emph{\bibinfo{journal}{APL Photonics}} \textbf{\bibinfo{volume}{7}}, \bibinfo{pages}{011301} (\bibinfo{year}{2022}).

\bibitem{zhelyeznyakov_large_2023}
\bibinfo{author}{Zhelyeznyakov, M.} \emph{et~al.}
\newblock \bibinfo{title}{Large area optimization of meta-lens via data-free machine learning}.
\newblock \emph{\bibinfo{journal}{Commun. Eng.}} \textbf{\bibinfo{volume}{2}}, \bibinfo{pages}{60} (\bibinfo{year}{2023}).

\bibitem{zhou_minimum_2015}
\bibinfo{author}{Zhou, M.}, \bibinfo{author}{Lazarov, B.~S.}, \bibinfo{author}{Wang, F.} \& \bibinfo{author}{Sigmund, O.}
\newblock \bibinfo{title}{Minimum length scale in topology optimization by geometric constraints}.
\newblock \emph{\bibinfo{journal}{Comput. Methods Appl. Mech. Eng.}} \textbf{\bibinfo{volume}{293}}, \bibinfo{pages}{266--282} (\bibinfo{year}{2015}).

\bibitem{hammond_photonic_2021}
\bibinfo{author}{Hammond, A.~M.}, \bibinfo{author}{Oskooi, A.}, \bibinfo{author}{Johnson, S.~G.} \& \bibinfo{author}{Ralph, S.~E.}
\newblock \bibinfo{title}{Photonic topology optimization with semiconductor-foundry design-rule constraints}.
\newblock \emph{\bibinfo{journal}{Opt. Express}} \textbf{\bibinfo{volume}{29}}, \bibinfo{pages}{23916} (\bibinfo{year}{2021}).

\bibitem{bulusu_generalization_2021}
\bibinfo{author}{Bulusu, S.}, \bibinfo{author}{Favoni, M.}, \bibinfo{author}{Ipp, A.}, \bibinfo{author}{Müller, D.~I.} \& \bibinfo{author}{Schuh, D.}
\newblock \bibinfo{title}{Generalization capabilities of translationally equivariant neural networks}.
\newblock \emph{\bibinfo{journal}{Phys. Rev. D}} \textbf{\bibinfo{volume}{104}}, \bibinfo{pages}{074504} (\bibinfo{year}{2021}).

\bibitem{cohen_group_2016}
\bibinfo{author}{Cohen, T.} \& \bibinfo{author}{Welling, M.}
\newblock \bibinfo{title}{Group equivariant convolutional networks}.
\newblock In \emph{\bibinfo{booktitle}{Proceedings of The 33rd International Conference on Machine Learning}}, vol.~\bibinfo{volume}{48}, \bibinfo{pages}{2990--2999} (\bibinfo{year}{2016}).

\bibitem{lilja_symmetry-informed_2026}
\bibinfo{author}{Lilja, V.~A.} \& \bibinfo{author}{Tassin, P.}
\newblock \bibinfo{title}{Symmetry-{{Informed Deep Learning}} for {{Electromagnetic Scattering}}}.
\newblock \bibinfo{howpublished}{Preprint at https://doi.org/10.48550/arXiv.2607.12810} (\bibinfo{year}{2026}).

\bibitem{elbek_tailoring_2026}
\bibinfo{author}{Elbek, P.}, \bibinfo{author}{Christiansen, R.~E.}, \bibinfo{author}{Aage, N.} \& \bibinfo{author}{Sigmund, O.}
\newblock \bibinfo{title}{Tailoring robust photonic components using stochastic topology optimization}.
\newblock \emph{\bibinfo{journal}{J. Opt. Soc. Am. B}} \textbf{\bibinfo{volume}{43}}, \bibinfo{pages}{1751--1763} (\bibinfo{year}{2026}).

\bibitem{mei_cascaded_2023}
\bibinfo{author}{Mei, F.} \emph{et~al.}
\newblock \bibinfo{title}{Cascaded metasurfaces for high-purity vortex generation}.
\newblock \emph{\bibinfo{journal}{Nat. Commun.}} \textbf{\bibinfo{volume}{14}}, \bibinfo{pages}{6410} (\bibinfo{year}{2023}).

\bibitem{hu_disordered_2023}
\bibinfo{author}{Hu, Z.}, \bibinfo{author}{Liu, C.} \& \bibinfo{author}{Li, G.}
\newblock \bibinfo{title}{Disordered optical metasurfaces: From light manipulation to energy harvesting}.
\newblock \emph{\bibinfo{journal}{Adv. Phys. X}} \textbf{\bibinfo{volume}{8}}, \bibinfo{pages}{2234136} (\bibinfo{year}{2023}).

\bibitem{zhou_optical_2024}
\bibinfo{author}{Zhou, H.} \emph{et~al.}
\newblock \bibinfo{title}{Optical computing metasurfaces: Applications and advances}.
\newblock \emph{\bibinfo{journal}{Nanophotonics}} \textbf{\bibinfo{volume}{13}}, \bibinfo{pages}{419--441} (\bibinfo{year}{2024}).

\bibitem{singh_active_2025}
\bibinfo{author}{Singh, S.}, \bibinfo{author}{Kumar, R.}, \bibinfo{author}{Singh, P.} \& \bibinfo{author}{Hegde, R.}
\newblock \bibinfo{title}{Active learning for efficient nanophotonics inverse design in large and diverse design spaces}.
\newblock \emph{\bibinfo{journal}{Opt. Express}} \textbf{\bibinfo{volume}{33}}, \bibinfo{pages}{20308--20321} (\bibinfo{year}{2025}).

\bibitem{wang_broadband_2017}
\bibinfo{author}{Wang, S.} \emph{et~al.}
\newblock \bibinfo{title}{Broadband achromatic optical metasurface devices}.
\newblock \emph{\bibinfo{journal}{Nat. Commun.}} \textbf{\bibinfo{volume}{8}}, \bibinfo{pages}{187} (\bibinfo{year}{2017}).

\bibitem{gopakumar_full-color_2024}
\bibinfo{author}{Gopakumar, M.} \emph{et~al.}
\newblock \bibinfo{title}{Full-colour 3d holographic augmented-reality displays with metasurface waveguides}.
\newblock \emph{\bibinfo{journal}{Nature}} \textbf{\bibinfo{volume}{629}}, \bibinfo{pages}{791--797} (\bibinfo{year}{2024}).

\bibitem{lee_metasurface_2018}
\bibinfo{author}{Lee, G.-Y.} \emph{et~al.}
\newblock \bibinfo{title}{Metasurface eyepiece for augmented reality}.
\newblock \emph{\bibinfo{journal}{Nat. Commun.}} \textbf{\bibinfo{volume}{9}}, \bibinfo{pages}{4562} (\bibinfo{year}{2018}).

\bibitem{cesa_program_2022}
\bibinfo{author}{Cesa, G.}, \bibinfo{author}{Lang, L.} \& \bibinfo{author}{Weiler, M.}
\newblock \bibinfo{title}{A program to build {E(N)}-equivariant steerable {CNN}s}.
\newblock In \emph{\bibinfo{booktitle}{Proceedings of the International Conference on Learning Representations (ICLR)}} (\bibinfo{year}{2022}).

\bibitem{akiba_optuna_2019}
\bibinfo{author}{Akiba, T.}, \bibinfo{author}{Sano, S.}, \bibinfo{author}{Yanase, T.}, \bibinfo{author}{Ohta, T.} \& \bibinfo{author}{Koyama, M.}
\newblock \bibinfo{title}{Optuna: A next-generation hyperparameter optimization framework}.
\newblock In \emph{\bibinfo{booktitle}{Proceedings of the 25th {ACM} {SIGKDD} International Conference on Knowledge Discovery and Data Mining}}, \bibinfo{pages}{2623--2631} (\bibinfo{year}{2019}).

\bibitem{kingma_adam_2015}
\bibinfo{author}{Kingma, D.~P.} \& \bibinfo{author}{Ba, J.}
\newblock \bibinfo{title}{Adam: A method for stochastic optimization}.
\newblock In \emph{\bibinfo{booktitle}{Proceedings of the International Conference on Learning Representations (ICLR)}} (\bibinfo{year}{2015}).

\end{thebibliography}

\section*{Acknowledgements}
We acknowledge partial financial support from Chalmers’ area of advance Nano (Excellence Grant) and from the Knut och Alice Wallenberg Stiftelse under Grant No.~2020.0119. Training data generation and NN training were performed on resources provided by the Swedish National Infrastructure for Computing (NAISS), at the Chalmers/C3SE and KTH/PDC sites, partially funded by the Swedish Research Council under Grant No.~2022-06725. The work was performed in part within the framework of the Excellence Center META-PIX.

\section*{Author contributions}
V.A.L. conceived the project, developed the computational methodology, and implemented the code.
All authors jointly conceptualized the experimental validation.
K.Z.K. developed the fabrication and measurement procedure and conducted experiments.
V.A.L. analyzed the data and all authors contributed to interpretation of the data.
V.A.L wrote the initial draft of the manuscript.
K.Z.K wrote the sections on fabrication and measurement.
P.T supervised and provided funding for the computational part of the project.
M.K. supervised and provided funding for the experimental verification.
All authors edited, reviewed, and approved the final manuscript.

% Begin extended figure numbering
\renewcommand{\figurename}{Extended Data Fig.}
\setcounter{figure}{0}

\newpage
\begin{figure}
    \centering
    \includegraphics[width=\linewidth]{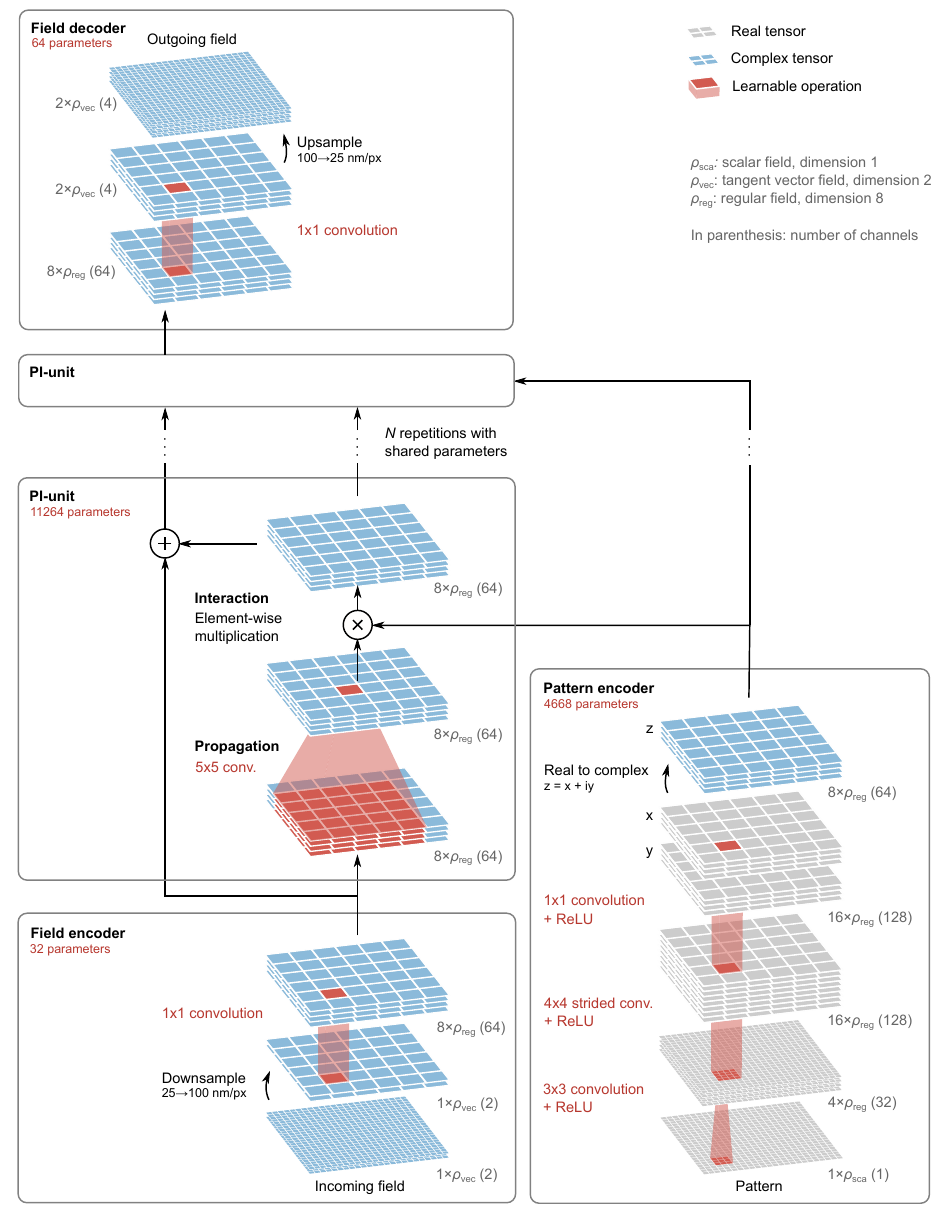}
    \caption{Detailed illustration of the EPIC-Net architecture. Field types are indicated next to each tensor; for example, ``$8\times\rho_\mathrm{reg}$ (64)'' means that the tensor transforms as eight copies of a regular feature field, comprising 64 channels in total \cite{cesa_program_2022}. All convolutions use periodic padding.}
    \label{fig:architecture}
\end{figure}

\newpage
\begin{figure}
    \centering
    \includegraphics[width=\linewidth]{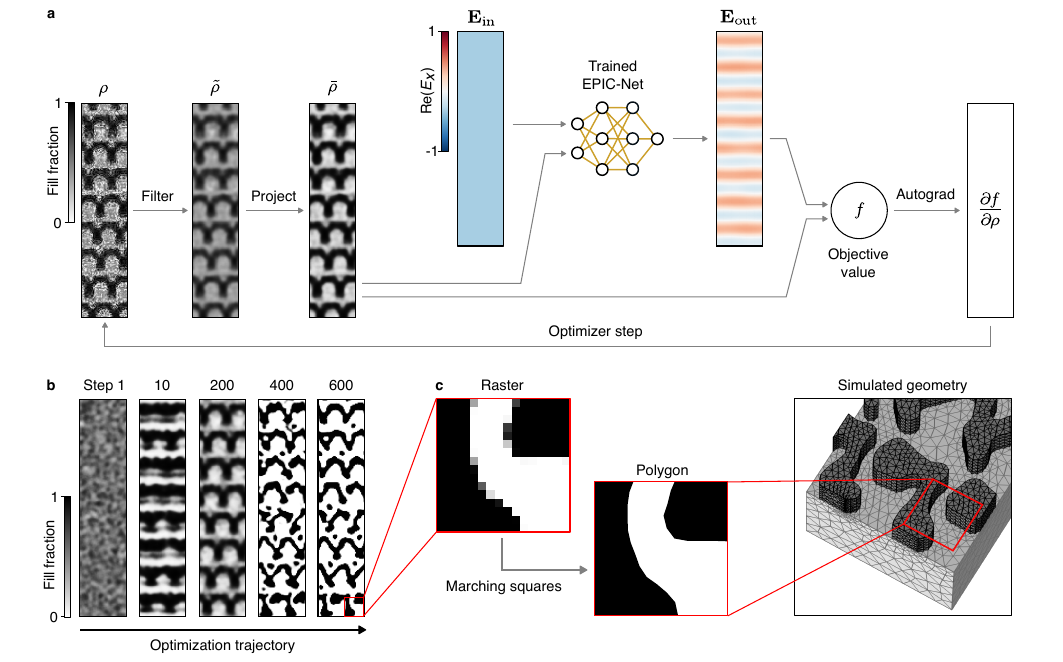}
    \caption{Inverse design by topology optimization. (a) Illustration of the topology optimization loop. (b) Evolution of the metasurface pattern during optimization. (c) Illustration of how the final metasurface pattern corresponds to the geometry simulated by the full-wave finite-element solver.}
    \label{fig:topopt}
\end{figure}

\newpage
\begin{figure}
    \centering
    \includegraphics[width=\linewidth]{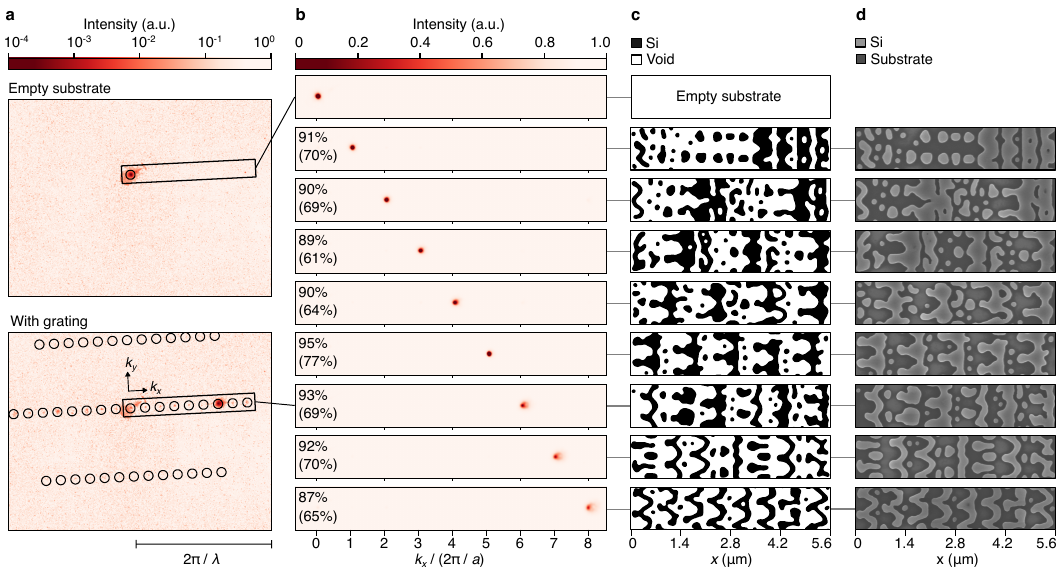}
    \caption{Grating geometries and efficiency calculation. 
    (a) Measured diffraction patterns for an empty substrate (top panel) and one of the metagratings (bottom panel). Rings in the lower panel show the expected positions of all diffraction orders allowed by the symmetry of the grating.
    (b) Zoomed-in measured diffraction patterns for all fabricated gratings. The calculated relative efficiency is shown in each panel. Absolute efficiencies are displayed in parentheses.
    (c) Topology-optimized grating geometries.
    (d) SEM images of the fabricated gratings.}
    \label{fig:grating_extra}
\end{figure}

\newpage
\begin{figure}
    \centering
    \includegraphics[width=\linewidth]{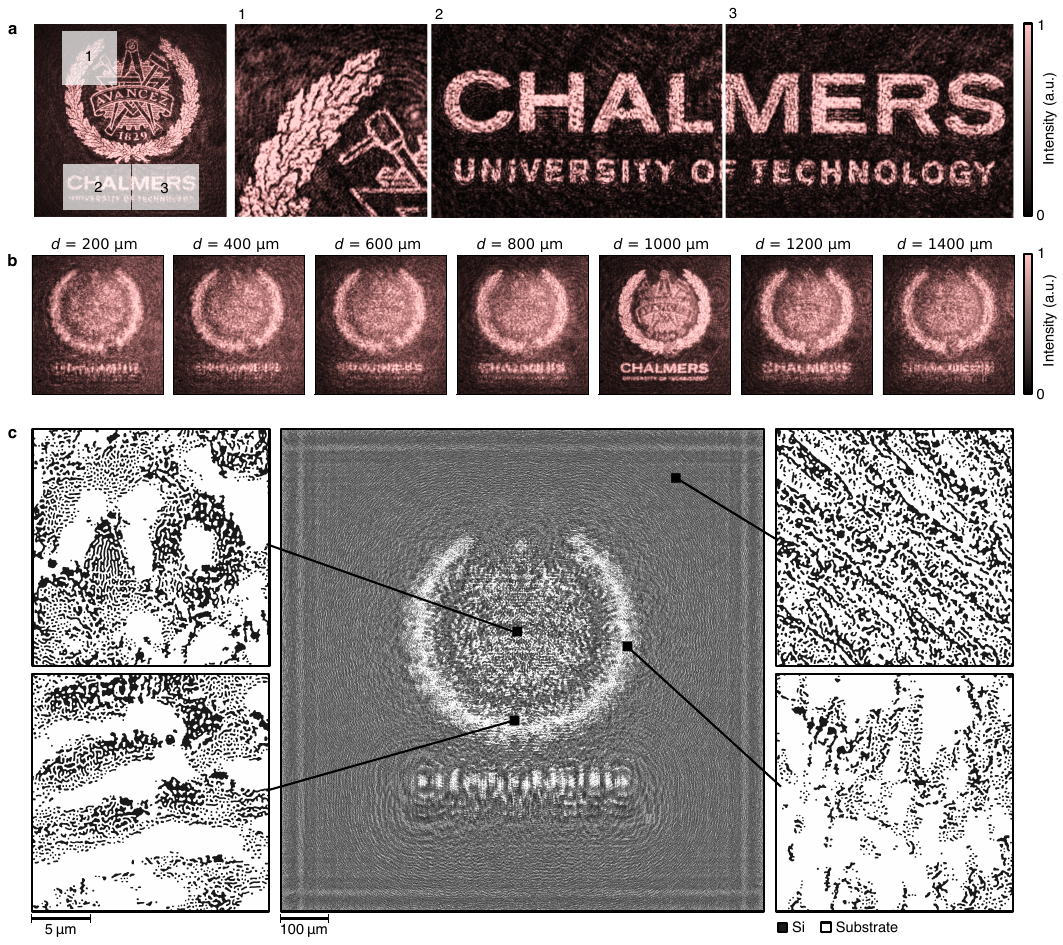}
    \caption{Additional results for the hologram. (a) Experimentally measured intensities with increased magnification at three different regions of the image plane. (b) Measured intensity distributions at different distances $d$ from the hologram. The target image appears at a distance of 1\,mm, as designed. (c) Zoomed-in images of various parts of the ideal pattern, demonstrating the diversity of features found by the optimization algorithm.}
    \label{fig:hologram_extra}
\end{figure}

\newpage
\begin{figure}
    \centering
    \includegraphics[width=\linewidth]{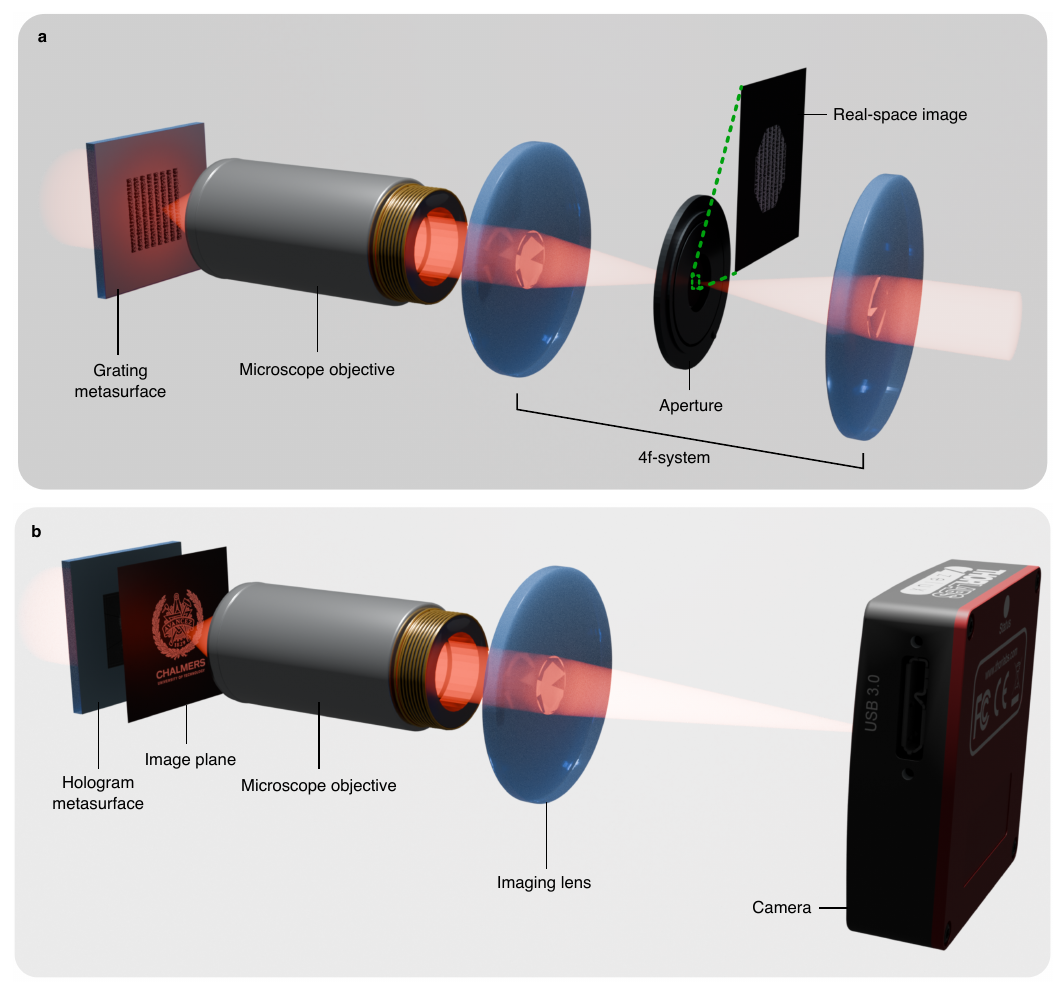}
    \caption{Illustration of the experimental setup. (a) Setup for the grating measurements. (b) Setup for the hologram measurements.}
    \label{fig:experimental_setup}
\end{figure}

\end{document}